\documentclass[10pt]{article}

\usepackage{hyperref}

\begin{document}
\title{Starting, Travelling \& Colliding Vortices: \\
DBD Plasma in Quiescent Air}
\author{Richard Whalley \& Kwing-So Choi \\
\\\vspace{6pt} Faculty of Engineering, University of Nottingham, Nottingham \\
NG7 2RD, UK}
\date{} 
\maketitle

\begin{abstract}
Development and interaction of starting vortices initiated by Dielectric Barrier Discharge (DBD) plasma actuators in quiescent air are illustrated in the attached fluid dynamics \href{http://hdl.handle.net/1813/13725}{video}. These include a series of smoke flow visualisations, showing the starting vortices moving parallel or normal to the wall at several different actuator configurations.   
\end{abstract}

\section{Introduction}
DBD plasma actuators consist of an upper and lower electrode separated by a thin dielectric material. On application of several kilovolts of ac power at kilohertz frequency between these electrodes, local ionization takes place around the upper electrode. This couples momentum to the surrounding fluid to induce a jet flow, which is caused by the movement of plasma ions to and from the dielectric surface. The charge build up on the dielectric surface opposes the charge of the exposed electrode, quenching the emission of the plasma discharge and stopping the plasma from collapsing into an arc. In other words, DBD plasma actuation is a self-limiting process \cite{Moreau2007}.\\
The attached fluid dynamics video examines the effects of the DBD plasma body force created at the actuator surface by showing a sequence of smoke flow visualisations in quiescent air. All images presented here have been captured using a high speed camera synchronised to a pulsed laser. On actuation of DBD plasma, a starting vortex \cite{Allen2007} is initiated which propagates along and away from the wall followed by a development of a laminar wall jet. When the plasma force is increased through an increase in the plasma power (from 20kHz to 25kHz at 5.50kV peak to peak), Kelvin-Helmholtz type instabilities are developed. Also illustrated in the video is the interaction of starting vortices from two opposing asymmetric electrodes. This leads to the collision of the starting vortices to develop into a wall-normal jet \cite{Santhanakrishnan2006}, with instabilities in the shear layer also visible.\\

\end{document}